\documentclass[12pt]{article}
\usepackage{graphicx}
\usepackage{mathrsfs}
\newcommand{\lbar}{\lambda\hspace{-0.45em}\raisebox{0.7ex}{-}\hspace{0.2em}}
\def\d{{\mathrm{d}}}
\def\ie{{\emph{i.e.}}}
\begin{document}
\title{\sc A Primer to Slow Light}
\author{Ulf Leonhardt\\
School of Physics and Astronomy\\
University of St Andrews\\
North Haugh, St Andrews, KY16 9SS, Scotland}

\maketitle

\begin{abstract}
Laboratory-based optical analogs of astronomical objects such as
black holes rely on the creation of light with an extremely low
or even vanishing group velocity (slow light). These brief notes
represent a pedagogical attempt towards elucidating this
extraordinary form of light. This paper is a contribution to the
book {\it Artificial Black Holes} edited by Mario Novello, Matt
Visser and Grigori Volovik. The paper is intended as a primer, an
introduction to the subject for non-experts, not as a detailed
literature review.
\end{abstract}

\newpage

%------------------------------------------------------
\section{Motivation}
%------------------------------------------------------
\index{slow light|(}

Creating a black hole with humble human resources is certainly a
fantastic idea, yet perhaps not entirely lunatic. Recent
experimental progress in quantum optics, in particular the
generation of slow light \cite{Leo:Slow,Leo:Liu,Leo:Philips} has
opened a route towards the formation of optical
event\index{optical event horizon} horizons for light or towards
other less dramatic but equally interesting phenomena.  Most of
the ideas have been discussed in a series of papers
\cite{Leo:Papers} and review articles \cite{Leo:Articles}. The
underlying atomic and optical physics is perhaps less familiar to
the audience of this book. Therefore, it might be worthwhile to
develop some of the physics behind slow light from basic
principles that are hopefully known to most readers.

%------------------------------------------------------
\section{Light-matter interaction}
%------------------------------------------------------

Consider an atomic medium capable of interacting with light, for
example a cell filled with Rubidium vapor or one of the alkali
Bose--Einstein\index{Bose--Einstein condensate} condensates. The
atoms that constitute the medium are electrically neutral but
polarizable by light. Since atoms are usually much smaller than
an optical wavelength, light experiences the atoms as dipoles (as
the lowest order in the multipole expansion of an electrically
neutral charge distribution \cite{Leo:Jackson}). In the dipole
approximation, the energy density of the light-matter interaction
is given by the negative scalar product $-{\bf P}\cdot{\bf E}$ of
the matter polarization ${\bf P}$ (dipole-moment density) and the
electric field ${\bf E}$ \cite{Leo:Jackson}. Therefore, the
Lagrangian of light and atomic matter reads in SI units
%%%%%%
\begin{equation}
\label{leo:l0} {\mathscr L} = \frac{\varepsilon_0}{2}\left(E^2 -
c^2 B^2\right) + {\bf P}\cdot{\bf E} + {\mathscr L}_A\,.
\end{equation}
%%%%%%
The first term describes the free electromagnetic field with $c$
denoting the speed of light in vacuum. The last term of the
Lagrangian characterizes the internal dynamics of the atoms. We
represent the electric field ${\bf E}$ and the magnetic field
${\bf B}$ in terms of the vector potential ${\bf A}$ in Coulomb
gauge \cite{Leo:Jackson}
%%%%%%
\begin{equation}
{\bf E} = -\partial_t {\bf A}\,,\quad {\bf B} = \nabla\times{\bf
A}\,,\quad\nabla\cdot{\bf A} = 0 \,.
\end{equation}
%%%%%%
Let us assume, for simplicity, that both the electromagnetic
field and the medium are uniform in two spatial directions in
Cartesian coordinates, but may vary in the third direction $z$.
Furthermore, we consider light with fixed polarization so that we
can concentrate on one component $A$ of the vector potential
${\bf A}$. The Lagrangian simplifies to
%%%%%%
\begin{equation}
{\mathscr L} = \frac{\varepsilon_0}{2}\left((\partial_t A)^2 - c^2
(\partial_z A)^2\right) - P\;\partial_t A + {\mathscr L}_A\,.
\end{equation}
%%%%%%
We obtain from the Euler--Lagrange equation the wave equation
%%%%%%
\begin{equation}
\label{leo:waveeq} \left(\partial_t^2-c^2\partial_z^2\right) E =
-\frac{1}{\varepsilon_0}\,\partial_t^2 P \,.
\end{equation}
%%%%%%
Atoms have a well-defined level structure such that light,
oscillating $10^{15}$ times per second, must match the atomic
transition frequencies, because otherwise the effect of ${\bf E}$
in $-{\bf P}\cdot{\bf E}$ is rapidly washed out. Given a certain
frequency range of light, the optical field thus interacts with a
selected few of the atomic levels, which greatly simplifies
matters.

Let us describe the atoms quantum-mechanically, while regarding
the electromagnetic field as classical. The simplest relevant
atomic system involves just two levels, say the ground state
$|\,a\,\rangle$ and the excited state $|\,b\,\rangle$. The
Hamiltonian of the two-level atom is simply
%%%%%%
\begin{eqnarray}
\hat{H} & = & \hbar{\omega_{ab}}\,|\,b\,\rangle\, \langle\,b\,| -
\frac{\kappa_{ab}}{2}\left(\hat{A} + \hat{A}^\dagger \right) E\,, \\
\hat{A} & = & |\,a\,\rangle\, \langle\,b\,|\,, \label{leo:aflip}
\end{eqnarray}
%%%%%%
where $\omega_{ab}$ denotes the atomic transition frequency and
$\kappa_{ab}$ corresponds to the atomic dipole moment, a real
number for transitions between bound states. In the Heisenberg
picture the transition operator $\hat{A}$ oscillates with
positive frequencies near $\omega_{ab}$. Therefore, $\hat{A}$
couples entirely to the negative-frequency component $E^{(-)}$ of
the electric field, while $\hat{A}^\dagger$ couples to the
positive-frequency component $E^{(+)}$. We arrive at the
effective Hamiltonian
%%%%%%
\begin{equation}
\label{leo:two}\hat{H} = \hbar{\omega_{ab}}\,|\,b\,\rangle\,
\langle\,b\,| - \frac{\kappa_{ab}}{2}\left(\hat{A}\,E^{(-)} +
\hat{A}^\dagger E^{(+)} \right) \,.
\end{equation}
%%%%%%
To describe the quantum state of the atom, we employ a density
matrix $\hat{\rho}$ that characterizes a statistical ensemble of
pure states $|\,\psi_a\,\rangle$ with probabilities $p_a$,
%%%%%%
\begin{equation}
\hat{\rho} = \sum_a p_a\,
|\,\psi_a\,\rangle\,\langle\,\psi_a\,|\,\,.
\end{equation}
%%%%%%
Probabilities are non-negative and sum up to unity. Consequently,
the density matrix has non-negative eigenvalues and is normalized
as ${\rm tr}\hat{\rho}=1$. In the Schr\"odinger picture, the
density matrix evolves while the operators are invariant in
time.  According to Lindblad's\index{Lindblad's theorem} theorem
\cite{Leo:Lindblad}, the evolution of a normalized and
non-negative density matrix is governed by the
master\index{master equation} equation \cite{Leo:Gardiner}
%%%%%%
\begin{equation}
\label{leo:master} \frac{\d\hat{\rho}}{\d t} =
\frac{i}{\hbar}\,[\hat{\rho},\hat{H}] - \sum_l
\gamma_l\left(\hat{A}_l^\dagger\;\hat{A}_l\;\hat{\rho} -
2\hat{A}_l\;\hat{\rho}\;\hat{A}_l^\dagger +
\hat{\rho}\;\hat{A}_l^\dagger\;\hat{A}_l \right) \,.
\end{equation}
%%%%%%
The Lindblad\index{Lindblad operator} operators $\hat{A}_l$
describe dissipative processes occurring at the rates $\gamma_l$,
for example the spontaneous emission from the excited state to
the ground state. As a result of the light-matter interaction, a
medium of $n_A$ atoms per volume generates a matter polarization
of
%%%%%%
\begin{equation}
\label{leo:pol} P = n_A \,\frac{\kappa_{ab}}{2}{\rm
tr}\left\{\hat{\rho}\left(\hat{A} + \hat{A}^\dagger\right)\right\}
= n_A \kappa_{ab} \; {\rm
Re}\,\langle\,a\,|\,\hat{\rho}\,|\,b\,\rangle \,.
\end{equation}
In this way the response of the atoms to the electric field
modifies the propagation of light given by the wave
equation~(\ref{leo:waveeq}).

Consider a medium at rest in a regime of linear response. Here the
medium integrates the local history of the electric field via the
susceptibility $\chi$,
%%%%%%
\begin{equation}
\label{leo:p} P = \varepsilon_0 \int_{-\infty}^{+\infty}
\chi(t-t')\;E(t')\;\d t' \,.
\end{equation}
%%%%%%
Causality implies that $P$ must not depend on the future of the
$E$ field, which restricts the integral (\ref{leo:p}) to the time
interval $(-\infty, 0]$ by requiring
%%%%%%
\begin{equation}
\chi(t) = 0 \quad {\rm for}\quad t<0 \,.
\end{equation}
%%%%%%
Consider the Fourier-transformed (spectral) susceptibility
%%%%%%
\begin{equation}
\label{leo:fourier} \tilde\chi(\omega) =
\frac{1}{2\pi}\int_{-\infty}^{+\infty} \chi(t)\;e^{i\omega t}
\;\d t = \frac{1}{2\pi}\int_{0}^{+\infty} \chi(t)\;e^{i\omega
t}\; \d t \,.
\end{equation}
%%%%%%
Because $\chi(t)$ is real, we get
%%%%%%
\begin{equation}
\label{leo:real} \tilde{\chi}(-\omega) = \tilde{\chi}^*(\omega)
\,.
\end{equation}
%%%%%%
Let us regard $\tilde{\chi}(\omega)$ as a function of complex
frequency $\omega$. When the imaginary part of $\omega$ is
positive, $\tilde\chi(\omega)$ cannot have singularities, because
here the Fourier integral contains a factor $\exp(-[{\rm
Im}\,\omega] t)$ that enforces convergence. Therefore,
$\tilde\chi(\omega)$ is analytic on the upper half plane.
Causality thus implies analyticity \cite{Leo:Nuss}. For a
non-dispersive medium $\tilde\chi(\omega)$ is constant over the
relevant frequency range and $\chi(t)$ is reduced to a delta
function, describing an instant response of the medium. In
dispersive media, $\tilde\chi(\omega)$ varies and the poles of
$\tilde\chi(\omega)$ on the lower half plane correspond to atomic
resonances. At the real $\omega$ axis the real part of
$\tilde{\chi}$ describes the dispersive properties of the medium,
whereas the imaginary part accounts for dissipation. Given an
analytic function $\tilde\chi(\omega)$ on the upper half plane
that decays sufficiently fast when $\omega\rightarrow\infty$, the
real and imaginary parts of $\tilde\chi(\omega)$ at the real
$\omega$ axis are related to each other by Hilbert transformations
\cite{Leo:Ablowitz} (Kramers-Kronig relations \cite{Leo:Nuss}).
The imaginary part of $\tilde\chi(\omega)$ is thus uniquely
determined by the real part and vice versa.

Dispersion influences the group velocity of light pulses. Suppose
that the medium properties do not vary significantly within the
scale of an optical wavelength. In this case we can characterize
completely the propagation of light pulses by the dispersion
relation between the wave number $k$ and the frequency $\omega$,
%%%%%%
\begin{equation}
k^2 = \frac{\omega^2}{c^2}\,\left[1+\tilde\chi(\omega,z)\right]
\,,
\end{equation}
%%%%%%
derived from the wave equation by Fourier transformation with
respect to space and time. A light pulse propagates like a
particle with Hamiltonian $\omega$ and momentum $k$, subject to
Hamilton's equations
%%%%%%
\begin{equation}
\label{leo:hamilton} \frac{\d k}{\d t} =
-\frac{\partial\omega}{\partial z}\,, \quad \frac{\d z}{\d t} =
\frac{\partial\omega}{\partial k} \,.
\end{equation}
%%%%%%
The group velocity $v_g$ is the velocity $\d z/\d t$ of the
fictitious light particle,
%%%%%%
\begin{equation}
v_g = \frac{\partial\omega}{\partial k} = \left(\frac{\partial k
}{\partial \omega}\right)^{-1} = \frac{\displaystyle
c}{\displaystyle n +
\frac{\omega}{2n}\,\frac{\partial\tilde{\chi}}{\partial\omega}}
\,, \qquad n = \sqrt{1 + \tilde{\chi}}\,. \label{leo:group}
\end{equation}
%%%%%%
Slow light \cite{Leo:Slow} involves an extremely dispersive medium
where $\partial\tilde{\chi}/\partial\omega$ is large and,
consequently, where the group velocity is very low (a few meters
per second, or lower).

%------------------------------------------------------
\section{Ordinary media}
%------------------------------------------------------

Before we embark on discussing extremely dispersive media, let us
consider an ordinary medium composed of two-level atoms at rest.
Assume that the atoms are identical, with equal transition
frequencies $\omega_{ab}$, and that they are affected by
dissipative relaxation processes. The dissipation transfers
excitations from the excited to the ground states, described by
the Lindblad\index{Lindblad operator} operator (\ref{leo:aflip}).
Assume that the transition rate $\gamma$ dominates the time scale
of the light-atom interaction. In this regime, relaxation forces
the atomic dipoles to follow the fields. In the case of linear
response we can decompose the electric field into Fourier
components. To analyse the response, it is sufficient to study
the reaction of one of the atoms to a single monochromatic wave
with frequency $\omega$. We characterize the field-strength of
the wave in terms of the Rabi\index{Rabi frequency} frequency
$\Omega$ with
%%%%%%
\begin{equation}
\Omega\; e^{-i\omega t} = \frac{\kappa_{ab}}{\hbar}\, E^{(+)}\,.
\end{equation}
%%%%%%
In the absence of relaxation, an atom would oscillate between the
ground and the excited state with frequency $\Omega$ (Rabi
flopping). On the other hand, relaxation leads to a stationary
state. To describe the stationary regime we use an appropriate
interaction picture.

\begin{figure}
%\label{F:Ulf1}
\begin{center}
\includegraphics[width=9cm]{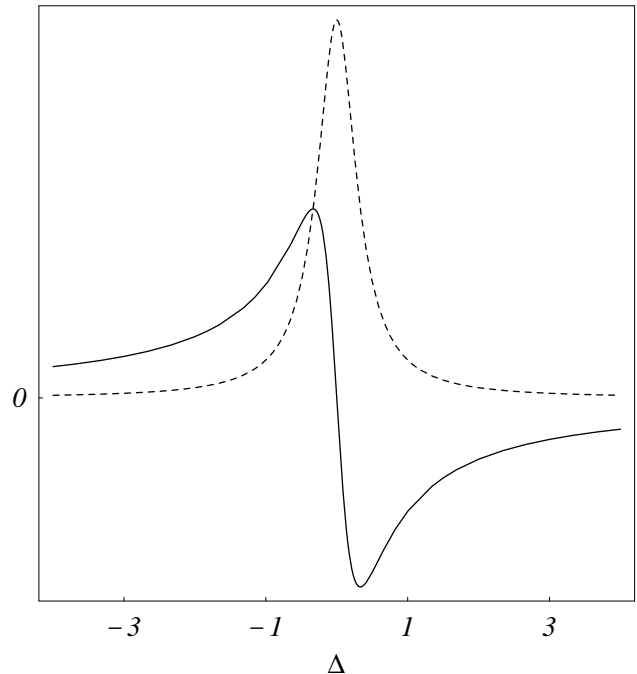}
\caption[Spectral susceptibility: ordinary dielectric medium.]
{\sl Spectral susceptibility of light in an ordinary dielectric
medium. The figure shows the real part (solid line) and the
imaginary part (dashed line) of $\tilde{\chi}(\omega)$ as a
function of the detuning $\Delta=\omega-\omega_0$ in arbitrary
units. The function is given by equations (\ref{leo:lorentz}) and
(\ref{leo:chi}). The line width $\gamma$ was set to $1/3$ in the
units used. The imaginary part of $\tilde{\chi}(\omega)$ has a
peak at the atomic resonance frequency (for zero detuning).
Outside the resonance the real part grows monotonically,
corresponding, according to equation (\ref{leo:group}), to a
positive group velocity. Near the resonance the dispersion
reaches a maximum. At the resonance the real part of
$\tilde{\chi}(\omega)$ decreases sharply (anomalous dispersion
leading to a negative group velocity). However, the interesting
spectral region of low or negative group velocity is totally
overshadowed by absorption.} \label{leo:ordinary}
\end{center}
\end{figure}

In an interaction picture, indicated by tildes over operators, the
dynamics with respect to a partial Hamiltonian $\hat{H}_0$ is
separated from the total evolution of the density matrix,
%%%%%%
\begin{equation}
\tilde{\rho} = \hat{U}_0^\dagger \;\hat{\rho}\; \hat{U}_0\,, \quad
\hat{U}_0=\exp\left(-\frac{i}{\hbar}\,\hat{H}_0\, t\right)\,.
\end{equation}
%%%%%%
To derive the evolution equation, we differentiate $\tilde{\rho}$
with respect to $t$ and apply the master equation
(\ref{leo:master}). Suppose that the commutator between
$\hat{A}_l$ and $\hat{H}_0$ is proportional to $\hat{A}_l$,
%%%%%%
\begin{equation}
\label{leo:flip} [\hat{A}_l,\hat{H}_0] = \hbar\omega_l\,\hat{A}_l
\,.
\end{equation}
%%%%%%
In this case $\hat{U}_0^\dagger \hat{A}_l\,\hat{U}_0$ gives
$\hat{A}_l\exp(-i\omega_l t)$, and thus the dissipative part of
the master equation (\ref{leo:master}) remains the same in the
interaction picture. The Hamiltonian is transformed according to
%%%%%%
\begin{equation}
\tilde{H} = \hat{U}_0^\dagger \;\hat{H}\;\hat{U}_0 - \hat{H_0} \,,
\end{equation}
%%%%%%
such that we obtain
%%%%%%
\begin{equation}
\label{leo:intmaster} \frac{\d\tilde{\rho}}{\d t} =
\frac{i}{\hbar}\,[\tilde{\rho},\tilde{H}] - \sum_l
\gamma_l\left(\hat{A}_l^\dagger\;\hat{A}_l\;\tilde{\rho} -
2\hat{A}_l\;\tilde{\rho}\;\hat{A}_l^\dagger +
\tilde{\rho}\;\hat{A}_l^\dagger\;\hat{A}_l \right) \,.
\end{equation}
%%%%%%
Returning to the two-level atom, we use the interaction picture
with respect to
%%%%%%
\begin{equation}
\hat{H}_0 = \hbar{\omega}\,|\,b\,\rangle\, \langle\,b\,|
\end{equation}
%%%%%%
that preserves the dissipative dynamics and leads to the
time-independent Hamiltonian
%%%%%%
\begin{equation}
\tilde{H} = \hbar({\omega_{ab}}-\omega)\;
|\,b\,\rangle\,\langle\,b\,| -
\textstyle{\frac{1}{2}}\left(\hat{A}\,\Omega^* + \hat{A}^\dagger
\Omega \right) \,.
\end{equation}
%%%%%%
The atom reaches a stationary state when the relaxation balances
the optical transition,
%%%%%%
\begin{equation}
\label{leo:stat2} \frac{i}{\hbar}\,[\tilde{\rho},\tilde{H}] =
\gamma\left(\hat{A}^\dagger\;\hat{A}\;\tilde{\rho} -
2\hat{A}\;\tilde{\rho}\;\hat{A}^\dagger +
\tilde{\rho}\;\hat{A}^\dagger\;\hat{A} \right) \,.
\end{equation}
%%%%%%
Given the normalization ${\rm tr}\hat{\rho}=1$, we can easily
solve the linear equation (\ref{leo:stat2}) for the density-matrix
components. Assuming linear response, we linearize the solution
$\tilde{\rho}$ in the Rabi frequency $\Omega$ and obtain
%%%%%%
\begin{equation}
\tilde{\rho} = \left[
\begin{array}{cc}
1 & \displaystyle{-\frac{\Omega^*}{2(\omega -
\omega_{ab}-i\gamma)}} \\
\displaystyle{-\frac{\Omega}{2(\omega - \omega_{ab}+i\gamma)}} & 0
\end{array}
\right] \,.
\end{equation}
%%%%%%
The positive-frequency component of the matter polarization
(\ref{leo:pol}) is therefore
%%%%%%
\begin{equation}
P^{(+)} = \frac{n_A}{2}\,\kappa_{ab}\,\langle\,b\,|\,
\hat{U}_0\,\tilde{\rho}\,\hat{U}_0^\dagger \,|\,a\,\rangle =
\frac{n_A}{4}\,\kappa_{ab}\,\tau_{ab}\,\Omega\,e^{-i\omega t}
\end{equation}
%%%%%%
where we have introduced
%%%%%%
\begin{equation}
\label{leo:lorentz} \tau_{ab} = -\frac{1}{\omega -
\omega_{ab}+i\gamma} \,,\quad \tau_{ba} = -\frac{1}{\omega +
\omega_{ab}+i\gamma} \,.
\end{equation}
%%%%%
For monochromatic light, $P^{(+)}$ gives simply $\varepsilon_0
\tilde{\chi}\,E^{(+)}$. Considering the property (\ref{leo:real})
of the spectral susceptibility we obtain the Lorentzian
%%%%%%
\begin{equation}
\label{leo:chi} \tilde{\chi} =
\frac{n_A}{4}\,\frac{\kappa_{ab}^2}{\varepsilon_0\hbar}\,
\left(\tau_{ab}+\tau_{ba}\right) \,.
\end{equation}
%%%%%%
Transforming to a real-time susceptibility $\chi(t)$ we see easily
that $\chi(t)$ responds within the relaxation time $\gamma$. In
accordance with causality, the Fourier-trans\-form\-ed
susceptibility $\tilde{\chi}$ is analytic on the upper half
plane. The single poles at $\pm\omega_{ab}-i\gamma$ correspond to
the atomic two-level resonance. On the real frequency axis,
$\tilde{\chi}$ is peaked at $\pm\omega_{ab}$ with the spectral
line width $\gamma$. Figure~\ref{leo:ordinary} illustrates how
the real and the imaginary part of the spectral susceptibility
depend on the frequency. The medium is most dispersive near the
resonance frequency $\omega_{ab}$ where, unfortunately, it is
also most absorptive. Instead of slowing down light, the medium
turns completely opaque.

%------------------------------------------------------
%------------------------------------------------------

%------------------------------------------------------
\section{Electromagnetically-Induced Transparency}
%------------------------------------------------------
\index{EIT|(}

Electromagnetically-Induced Transparency (EIT) \cite{Leo:EIT} has
served as a method to slow down light significantly
\cite{Leo:Slow} or, ultimately, to freeze light completely
\cite{Leo:Liu,Leo:Philips}. Like other successful techniques, EIT
is based on a simple idea \cite{Leo:EIT}: A control beam of laser
light couples the upper levels of an atom, and, in this way, the
beam strongly modifies the optical properties of the atom. In
particular, the coupling of the excited states affects the
transition from the atomic ground state to one of the upper
states, \ie, the ability of the atom to absorb probe photons with
matching transition frequency. Destructive quantum interference
between the paths of the transition process turns out to
eliminate absorption at exact resonance \cite{Leo:EIT}. A medium
composed of such optically-manipulated atoms is transparent in a
spectral region where it would otherwise be completely opaque. In
the vicinity of the transparency frequency $\omega_0$ the medium
is highly dispersive, \ie, the refractive index changes within a
narrow frequency interval. In turn, probe light pulses with a
carrier frequency at $\omega_0$ travel with a very low group
velocity $v_g$ \cite{Leo:Harris}.

%------------------------------------------------------
\begin{figure}
%\label{F:Ulf2}
\begin{center}
\includegraphics[width=9cm]{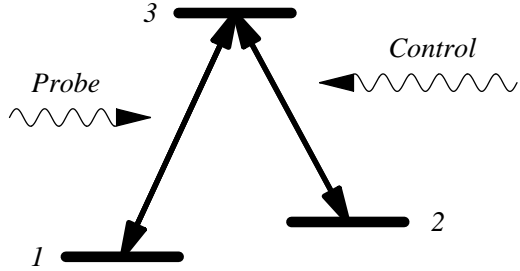}
\caption[Three-level atom: Electromagnetically Induced
Transparency.]  {\sl Three-level atom in a regime of
Electromagnetically-Induced Transparency. The control beam
couples the levels 2 and 3, which influences strongly the optical
properties of the atom for a weaker probe beam tuned to the
transition 1$\leftrightarrow$3. } \label{leo:levels}
\end{center}
\end{figure}
%------------------------------------------------------

Consider the three-level atom illustrated in
figure~\ref{leo:levels}. The atom is characterized by the
energy-level differences $\hbar\omega_{12}$ and
$\hbar\omega_{23}$ with $\omega_{12} + \omega_{23} = \omega_{13}
\equiv \omega_0$. Typically, the transition frequencies
$\omega_{13}$ and $\omega_{23}$ are in the optical range of the
spectrum or in the near infrared ($10^{15}{\rm Hz}$), whereas the
frequency $\omega_{12}$ is much lower ($10^{9}{\rm Hz}$). The
atom is subject to fast relaxation mechanisms ($10^{6}{\rm Hz}$)
that transport atomic excitations from the $|\,3\,\rangle$ state
down to $|\,1\,\rangle$ and from $|\,3\,\rangle$ to
$|\,2\,\rangle$, mainly caused by spontaneous emission. Hardly
any excitations move from  $|\,2\,\rangle$ to $|\,1\,\rangle$,
because the spontaneous emission rate is proportional to the cube
of the frequency \cite{Leo:Loudon}. Here the relaxation may be
dominated by other processes, for instance by spin-exchanging
collisions. Without relaxation the dynamics of the atom is
governed by the Hamiltonian
%%%%%%
\begin{equation}
\label{leo:h} \hat{H} = \left[
\begin{array}{ccc}
0 & 0 & -\frac{1}{2}\,\kappa_{13}\,E_p^{(-)} \\[3mm]
0 & \hbar\omega_{12} & -\frac{1}{2}\,\kappa_{23}\,E_c^{(-)}\\[3mm]
-\frac{1}{2}\,\kappa_{13}\,E_p^{(+)} &
-\frac{1}{2}\,\kappa_{23}\,E_c^{(+)} & \hbar\omega_{13}
\end{array}
\right] \,.
\end{equation}
%%%%%%
The Hamiltonian represents the atomic level structure and
describes the $-\hat{P}E$ interaction with light, considering here
only the frequency components $E_p^{(\pm)}$ and $E_c^{(\pm)}$
that match approximately the level structure. The $E_p$ and $E_c$
fields are the probe and control light respectively. We describe
relaxation phenomenologically by the flip processes
%%%%%%
\begin{equation}
\label{leo:relax} \hat{A}_1 =
|\,1\,\rangle\,\langle\,3\,|\,\,,\quad \hat{A}_2 =
|\,2\,\rangle\,\langle\,3\,|\,\,,
\end{equation}
%%%%%%
occurring at the rates $\gamma_1$ and $\gamma_2$, typically a few
$10^{6}\,{\rm Hz}$. Suppose that the three-level atom is
illuminated with monochromatic control light at frequency
$\omega_c$ in exact resonance with the $2 \leftrightarrow 3$
transition,
%%%%%%
\begin{equation}
\label{leo:res} \omega_c = \omega_{23} \,.
\end{equation}
%%%%%%
Consider a regime of linear response. In this case we can
decompose the probe field into monochromatic waves, to describe
completely the reaction of the atom. We characterized the two
light fields involved by their Rabi frequencies $\Omega_c$ and
$\Omega_p$, defined as
%%%%%%
\begin{equation}
\Omega_c\, e^{-i\omega_c t} = \frac{\kappa_{23}}{\hbar}\,
E_c^{(+)} \,,\quad \Omega_p\, e^{-i\omega t} =
\frac{\kappa_{13}}{\hbar}\, E_p^{(+)} \,.
\end{equation}
%%%%%%
The Rabi frequencies set the time scales of atomic transitions
caused by the applied light fields. The control beam shall
dominate all processes,
%%%%%%
\begin{equation}
\label{leo:r1} |\,\Omega_c| \gg
|\,\Omega_p|\,,\,\gamma_1\,,\,\gamma_2 \,.
\end{equation}
%%%%%%
Mediated by relaxation, the atomic dipoles lose any initial
oscillations they might have possessed and follow the optical
fields. To find the stationary state, we utilize an interaction
picture generated by
%%%%%%
\begin{equation}
\hat{H}_0 = \hbar \left[
\begin{array}{ccc}
0 & 0 & 0\\
0 & \omega-\omega_c & 0\\
0 & 0 & \omega
\end{array}
\right] \,.
\end{equation}
%%%%%%
Due to commutation relations of the type (\ref{leo:flip}) the
dissipative part of the master equation is not changed in the
interaction picture. The transformed Hamiltonian has become
time-independent,
%%%%%%
\begin{equation}
\label{leo:hstat} \tilde{H} = -\hbar \left[
\begin{array}{ccc}
0 & 0 & \frac{1}{2}\,\Omega_p^* \\[3mm]
0 & \omega-\omega_0 & \frac{1}{2}\,\Omega_c^*\\[3mm]
\frac{1}{2}\,\Omega_p & \frac{1}{2}\,\Omega_c & \omega-\omega_0
\end{array}
\right] \,.
\end{equation}
%%%%%%
Similar to a two-level atom in a stationary state
(\ref{leo:stat2}), the optical transitions should balance the
relaxation processes,
%%%%%%
\begin{equation}
\label{leo:stat}  \frac{i}{\hbar}\,[\tilde{\rho},\tilde{H}] =
\sum_{l=1}^2 \gamma_l\left(\hat{A}_l^\dagger\hat{A}_l\tilde{\rho}
- 2\hat{A}_l\tilde{\rho}\hat{A}_l^\dagger +
\tilde{\rho}\hat{A}_l^\dagger\hat{A}_l \right) \,.
\end{equation}
%%%%%%
We could solve exactly the linear equation (\ref{leo:stat}) for
the matrix elements of $\tilde{\rho}$ with ${\rm
tr}\tilde{\rho}=1$ (using computerized formula manipulation, for
example), but without gaining much insight. Fortunately, since we
are interested in the regime (\ref{leo:r1}), we can find
transparent approximations. Suppose first that also
%%%%%%
\begin{equation}
\label{leo:r2} |\,\Omega_c| \gg |\,\omega-\omega_0 |\,.
\end{equation}
%%%%%%
We expand the solution of equation\ (\ref{leo:stat}) to quadratic
order in the small quantities (\ref{leo:r1}) and (\ref{leo:r2}),
and get
%%%%%%
\begin{equation}
\tilde{\rho} = \left[
\begin{array}{ccc}
{\displaystyle 1 - \frac{|\,\Omega_p|^2}{|\,\Omega_c|^2}} &
{\displaystyle -\frac{\Omega_p^*}{\Omega_c^*}} & {\displaystyle
\frac{2(\omega-\omega_0)}{|\,\Omega_c|^2}\,\Omega_p^*}\\[4mm]
{\displaystyle -\frac{\Omega_p}{\Omega_c}} & {\displaystyle
\frac{|\,\Omega_p|^2}{|\,\Omega_c|^2}} & 0\\[4mm]
{\displaystyle
\frac{2(\omega-\omega_0)}{|\,\Omega_c|^2}\,\Omega_p} & 0 & 0
\end{array}
\right] \,. \label{leo:stationary}
\end{equation}
%%%%%%
We proceed similarly to our analysis of the two-level atom and
find, in the positive-frequency range, the spectral susceptibility
%%%%%%
\begin{equation}
\tilde{\chi} = \frac{2\alpha}{\omega_0}\, (\omega - \omega_0) \,,
\label{leo:linear}
\end{equation}
%%%%%%
given here in terms of the parameter
%%%%%%
\begin{equation}
\alpha = \frac{n_A}{2}\,
\frac{\kappa_{13}^2}{\kappa_{12}^2}\,\frac{\hbar\omega_0}{\varepsilon_0\,
|E_c|^2}\,. \label{leo:alpha}
\end{equation}
%%%%%%
The spectral susceptibility $\tilde{\chi}$ depends linearly on the
detuning $\omega-\omega_0$ and vanishes at the resonance
frequency. Here the phase velocity of light is exactly the speed
of light in vacuum, $c$, but the group velocity (\ref{leo:group})
is reduced by $(1+\alpha)$,
%%%%%%
\begin{equation}
v_g = \frac{c}{1 + \alpha} \,. \label{leo:vg}
\end{equation}
%%%%%%
We call the $\alpha$ parameter (\ref{leo:alpha}) group index. The
parameter is proportional to the ratio between the probe-light
energy per photon, $\hbar\omega_0$, and the control-light energy
per atom, $\varepsilon_0\,|E_c|^2/ n_A$. Consequently, the less
intense the control beam is the slower the probe light is, a
paradoxical behavior. Taken to the extreme, the group velocity
would vanish when the control beam is totally dimmed. However, in
the stationary regime that we are considering, the control beam
should dominate (\ref{leo:r1}) and the detuning should be small
compared with the modulus of the control's Rabi frequency.

\begin{figure}
%\label{F:Ulf3}
\begin{center}
\includegraphics[width=9cm]{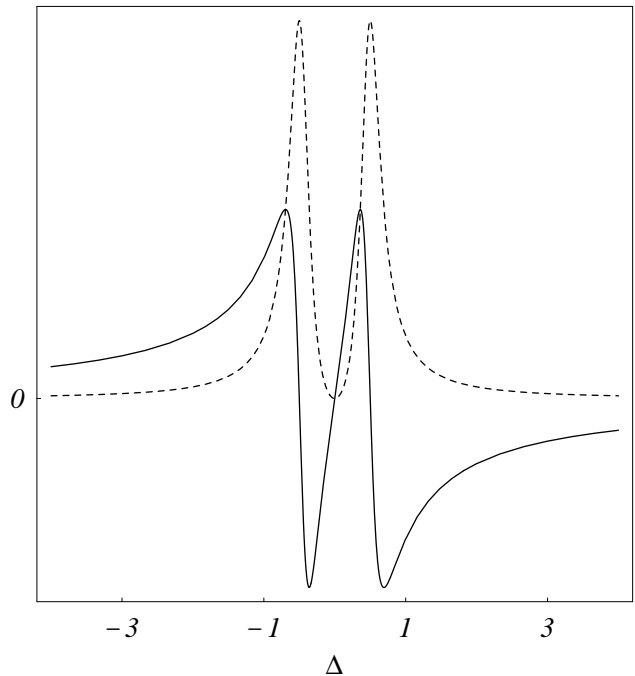}
\caption[Spectral susceptibility: Electromagnetically Induced
Transparency] {\sl Susceptibility of the probe light in a medium
with Electromagnetically-Induced Transparency. The figure shows
the real part (solid line) and the imaginary part (dashed line)
of the spectral susceptibility $\tilde{\chi}(\omega)$ as a
function of the detuning $\Delta=\omega-\omega_0$ in units of the
Rabi frequency $\Omega_c$ of the control beam, given by equations
(\ref{leo:chi}) and (\ref{leo:tau}). The line width
$\gamma_1+\gamma_2$ was set to $\Omega_c/3$. The parameters used
agree with the ones in figure~\ref{leo:ordinary}. Comparing the
two figures, we see that EIT radically alters the susceptibility
in a spectral region around the probe resonance $\omega_0$. Here
the imaginary part of $\tilde{\chi}(\omega)$ vanishes. The medium
has become transparent where it would be completely opaque
without the influence of the control beam. In the transparency
window the real part of $\tilde{\chi}(\omega)$ increases linearly
with a steep slope, indicating that the medium is extremely
dispersive.  As a consequence of equation (\ref{leo:group}) the
group velocity of the probe light is significantly reduced.}
\label{leo:susc}
\end{center}
\end{figure}

Apparently, the linear spectral slope (\ref{leo:linear}) of the
susceptibility ought to be limited. To find the limitation, we
expand the exact stationary state of the master equation
(\ref{leo:intmaster}) to linear order in $\Omega_p$, in
accordance with a regime of linear response. We obtain a spectral
susceptibility (\ref{leo:chi}) with
%%%%%%
\begin{equation}
\label{leo:tau} \tau_{13} =
-\frac{\omega-\omega_0}{(\omega-\omega_0)^2 +
i(\omega-\omega_0)(\gamma_1 + \gamma_2)
-\frac{1}{4}\,|\Omega_c|^2} \,.
\end{equation}
%%%%%%
One can easily verify that the poles of $\tilde{\chi}$ are
located on the lower half plane of the complex frequency $\omega$,
in agreement with the causality of the medium's response. We
expand formula (\ref{leo:tau}) in powers of
${(\omega-\omega_0)}/{|\,\Omega_c|}$ and see that the medium
becomes dissipative when the condition
%%%%%%
\begin{equation}
|\,\omega-\omega_0\,| \ll \frac{|\,\Omega_c|^2}{\gamma_1 +
\gamma_2}
\end{equation}
%%%%%%
is violated. For a large detuning we can ignore the
$\frac{1}{4}\,|\Omega_c|^2$ term in the susceptibility
(\ref{leo:tau}). We obtain the simple Lorentzian
(\ref{leo:lorentz}) of an ordinary medium, with the combined line
width $\gamma = \gamma_1 + \gamma_2$. Outside the narrow
transparency window of EIT, the absorption of the medium has even
slightly increased, because the control beam couples the medium
atoms to a second dissipative transition process. The maximally
tolerable detuning for transparency is proportional to the group
velocity, since $v_g$ is proportional to $|\,\Omega_c|^2$ for
sufficiently slow light. In practice the detuning is usually
limited by $\epsilon v_g \omega_0/c$ with $\epsilon$ in the order
of a few $10^{-3}$. The transparency window concerns slow light
in moving media, because of the Doppler effect. An atom with
velocity $u$ causes a Doppler detuning of $u\,\omega_0/c$. If we
fix the spectral range in the laboratory frame, the maximally
tolerable velocity is $\epsilon v_g$. EIT is velocity-selective.
Figure~\ref{leo:susc} illustrates the spectral susceptibility.

%------------------------------------------------------
%------------------------------------------------------
\index{EIT|)}
%------------------------------------------------------
\section{Dark-state dynamics}
%------------------------------------------------------
\index{dark state|(}
Suppose that a dominant and monochromatic control beam has, after
relaxation, prepared the atom in the stationary state
(\ref{leo:stationary}). How will the atom evolve when the control
and probe strengths vary \cite{Leo:Fleisch}?  First we note that
the state (\ref{leo:stationary}) is statistically pure [to
quadratic order in the small quantities (\ref{leo:r1}) and
(\ref{leo:r2})] so that
%%%%%%
\begin{equation}
\label{leo:purity} \rm{tr}\{\hat{\rho}^2\} = 1 \,.
\end{equation}
%%%%%%
When the purity condition (\ref{leo:purity}) is satisfied the
density matrix contains a single state vector \cite{Leo:Gardiner}
%%%%%%
\begin{equation}
\hat{\rho} = |\,\psi_0\,\rangle\,\langle\,\psi_0\,|
\end{equation}
%%%%%%
with, in our case,
%%%%%%
\begin{equation}
\label{leo:psi0} |\,\psi_0\,\rangle = \hat{U}_0\, N_0
\left(|\,1\,\rangle - \frac{\Omega_p}{\Omega_c}\,|\,2\,\rangle +
\frac{2(\omega-\omega_0)}{|\,\Omega_c|^2}\,\Omega_p\,|\,3\,\rangle
\right) \,.
\end{equation}
%%%%%%
The stationary state does not depend on the relaxation rates but
only on the parameters of the Hamiltonian (\ref{leo:hstat}).
Remarkably, even when the parameters vary, the state is protected
from further relaxation, as long as the $|\,3\,\rangle$ component
is small,
%%%%%%
\begin{equation}
\label{leo:condition} \rho_{33} = \langle \, 3\,
|\,\hat{\rho}\,|\, 3 \,\rangle \ll 1 \,.
\end{equation}
%%%%%%
Once the atom is in a pure state with sparsely populated top
level, the purity (\ref{leo:purity}) does not change during the
evolution (\ref{leo:master}),
%%%%%%
\begin{equation}
\d \rm{tr}\{\hat{\rho}^2\} = 2 \rm{tr}\{\hat{\rho}\,\d\hat{\rho}\}
= 4\left[\gamma_1(1-\rho_{11}) + \gamma_2(1-\rho_{22}) \right]
\rho_{33}\, \d t \,.
\end{equation}
%%%%%%
The pure state so protected is called a dark state
\cite{Leo:Arimondo}. Although dark states are initially prepared
due to the relaxation of the atomic dipoles, having so adapted to
the light fields, they are no longer prone to dissipation.

Suppose that the control and the probe strengths vary. How does a
dark state follow the light? In the case (\ref{leo:condition})
the state of the atom is dominated by its components in the
subspace spanned by the two lower levels $|\,1\,\rangle$ and
$|\,2\,\rangle$. If we find a state $|\,\psi\,\rangle$ that
describes correctly the dynamics (\ref{leo:master}) in this
subspace, the third component $\langle\,3\,|\,\psi\,\rangle$ must
be correct as well, to leading order in $\rho_{33}$. The lower
ranks enslave the top level. Since the relaxation processes
(\ref{leo:relax}) do not operate within the lower subspace, we
can ignore dissipation entirely, to find the dominant state of
the atom. We represent both control and probe light in terms of
variable Rabi\index{Rabi frequency} frequencies,
%%%%%%
\begin{equation}
\Omega_c\, e^{-i\omega_c t} = \frac{\kappa_{23}}{\hbar}\,
E_c^{(+)} \,,\quad \Omega_p\, e^{-i\omega_0 t} =
\frac{\kappa_{13}}{\hbar}\, E_p^{(+)} \,,
\end{equation}
%%%%%%
defined here with respect to the atomic transition frequencies
$\omega_c=\omega_{23}$ and $\omega_0=\omega_{13}$. We write down
the state vector
%%%%%%
\begin{equation}
\label{leo:dark} |\,\psi\,\rangle = \hat{U}_0 \, N
\left(|\,1\,\rangle - \frac{\Omega_p}{\Omega_c}\,|\,2\,\rangle +
\frac{2N_0^2}{\Omega_c^*}\,i\partial_t\,\frac{\Omega_p}{\Omega_c}
\,|\,3\,\rangle \right)
\end{equation}
%%%%%%
with the abbreviations
%%%%%%
\begin{eqnarray}
\hat{U}_0 &=& \left[
\begin{array}{ccc}
1 & 0 & 0 \\
0 & e^{-i\omega_{12}t} & 0\\
0 & 0 & e^{-i\omega_0t}
\end{array}\right],
\\[1mm]
\frac{\Omega_p}{\Omega_c} &=&
\left|\frac{\Omega_p}{\Omega_c}\right|\,e^{i\theta}\,,
\\[2mm]
N_0 &=& \left(1 +
\frac{|\,\Omega_p|^2}{|\,\Omega_c|^2}\right)^{-1/2} \,,
\\[2mm]
N &=&
N_0\,\exp\left(-i\int\frac{|\,\Omega_p|^2\,d\theta}{|\,\Omega_p|^2
+ |\,\Omega_c|^2}\right)\,.
\end{eqnarray}
%%%%%%
In a stationary regime under the conditions (\ref{leo:r1}) and
(\ref{leo:r2}) the vector (\ref{leo:dark}) agrees with the dark
state (\ref{leo:psi0}). We see from the properties
%%%%%%
\begin{equation}
\partial_t N = -N N_0^2\,\frac{\Omega_p^*}{\Omega_c^*}\,\partial_t\,
\frac{\Omega_p}{\Omega_c},\qquad
\partial_t N\,\frac{\Omega_p}{\Omega_c} = N N_0^2\,\partial_t\,
\frac{\Omega_p}{\Omega_c},
\end{equation}
%%%%%%
that $|\,\psi\,\rangle$ satisfies the differential equation
%%%%%%
\begin{equation}
\label{leo:dyn} i\hbar\,\partial_t\,|\,\psi\,\rangle =
\hat{H}\,|\,\psi\,\rangle + i\hbar\,\partial_t\,
\langle\,3\,|\,\psi\,\rangle |\,3\,\rangle \,.
\end{equation}
%%%%%%
Consequently, the vector (\ref{leo:dark}) describes correctly the
dynamics of the atom in the lower-level subspace. Therefore, the
atom remains in the dark state (\ref{leo:dark}), as long as the
atom's evolution never leads to an overpopulation at the top
level $|\,3\,\rangle$. The initial relaxation-dominated regime
has prepared the dark state, but later the atom follows
dynamically without relaxation \cite{Leo:Fleisch}.

We calculate the matter polarization (\ref{leo:pol}) generated by
the dark states of the atoms that constitute the medium. The
positive-frequency component of $P$ is
%%%%%%
\begin{eqnarray}
P^{(+)} & = & \frac{n_A}{2}\, \kappa_{31}\,\langle\,3\,
|\,\psi\,\rangle\langle\,\psi\,|\,1\,\rangle
\nonumber\\
& = & \frac{n_A}{2}\, \kappa_{31}\,e^{-i\omega_0 t} N_0^4\,
\frac{2}{\Omega_c^*}\,i\partial_t\,\frac{\Omega_p}{\Omega_c}
\nonumber\\
& = &
n_A\,\frac{\kappa_{31}^2}{\hbar}\,\frac{N_0^4}{|\,\Omega_c|^2}
\left(i\partial_t - \omega_0 -
i\frac{(\partial_t|\,\Omega_c|)}{|\,\Omega_c|} + \dot{\theta}_c
\right)E^{(+)}_p,
\end{eqnarray}
%%%%%%
with $\theta_c = {\rm arg}\Omega_c$. Assume, for simplicity, that
$\Omega_c$ is real. Otherwise we can easily incorporate the phase
$\theta_c$ of the control field in the phase of the electric
field without affecting the wave equation (\ref{leo:waveeq}), as
long as $\theta_c$ varies slowly compared with the optical
frequency $\omega_0$. We adopt the definition (\ref{leo:alpha})
of the group index $\alpha$, and get
%%%%%%
\begin{equation}
\frac{1}{\varepsilon_0}\,\partial_t^2 P^{(+)} \approx
-N_0^4\alpha\,2\omega_0\left(i\partial_t - \omega_0 -
i\frac{\dot{\alpha}}{2\alpha}\right)E^{(+)}_p \,.
\end{equation}
%%%%%%
We approximate
%%%%%%
\begin{equation}
2\omega_0(i\partial_t - \omega_0)\,E^{(+)}_p \approx (i\partial_t
+ \omega_0)(i\partial_t - \omega_0)\,E^{(+)}_p =
-(\partial_t^2+\omega_0^2)\,E^{(+)}_p,
\end{equation}
%%%%%%
and obtain from the general wave equation (\ref{leo:waveeq}) an
equation that is valid for both the positive and the negative
frequency component of the probe light,
%%%%%%
\begin{equation}
\label{leo:saturation} \left[\partial_t^2 - c^2 \partial_z^2 +
N_0^4\left(\partial_t \alpha\, \partial_t +
\alpha\,\omega_0^2\right)\right] E_p = 0 \,.
\end{equation}
%%%%%%
The dark-state dynamics may lead to a non-linear effect of the
medium, described by the $N_0^4$ factor in the wave equation
(\ref{leo:saturation}). However, when the probe is significantly
weaker than the control light, the medium responds linearly,
%%%%%%
\begin{equation}
\label{leo:wave} \left(\partial_t(1+\alpha)\partial_t - c^2
\partial_z^2 + \alpha\,\omega_0^2\right) E_p = 0 \,.
\end{equation}
%%%%%%
This wave equation governs the propagation of slow light in a
regime of linear response and undisturbed dark-state dynamics.

\index{dark state|)}
%------------------------------------------------------
\section{Slow-light pulses}
%------------------------------------------------------
\index{slow light!pulse|(}

Consider a pulse of probe light in an EIT\index{EIT} medium with
variable group index (\ref{leo:alpha}). Suppose that the group
velocity (\ref{leo:vg}) does not vary much over the scale of an
optical wavelength ($0.5\times10^{-6}{\rm m}$) or an optical cycle
($10^{-15}{\rm s}$). In this case we could apply the Hamiltonian
theory (\ref{leo:hamilton}) of a fictitious light particle to
predict the position of the pulse peak. Because particle
trajectories must not split, a slowly varying group index cannot
cause reflection. Suppose that the pulse is traveling to the
right. Then the pulse will continue to do so, and we can express
the slow-light wave as
%%%%%%
\begin{equation}
\label{ansatz} E_p = {\cal E} \exp(ikz-i\omega t) + {\cal E}^*
\exp(-ikz+i\omega t)\,,\quad k = \frac{\omega}{c}\,,
\end{equation}
%%%%%%
assuming that the electric-field amplitude ${\cal E}$ is slowly
varying compared with the rapid optical oscillations. We
approximate
%%%%%%
\begin{eqnarray}
\exp(-ikz+i\omega t)\, \partial_t^2 E_p^{(+)} &\approx&
\left(-\omega^2
- 2i\omega\partial_t \right) {\cal E}\,,\nonumber\\
\exp(-ikz+i\omega t)\, \partial_t E_p^{(+)}\, &\approx& -i\omega
{\cal E} \,,
\nonumber\\
\exp(-ikz+i\omega t)\, \partial_z^2 E_p^{(+)} &\approx& \left(-k^2
+ 2ik\partial_t \right) {\cal E}\,.
\end{eqnarray}
%%%%%%
and get from the wave equation (\ref{leo:wave})
%%%%%%
\begin{eqnarray}
-2i\omega(1+\alpha)\,\partial_t {\cal E} &\approx& \left((1 +
\alpha)\omega^2 + i\omega\dot{\alpha} - c^2\,k^2 +
2ikc^2\partial_z - \alpha\omega_0^2 \right) {\cal E}
\nonumber\\
&=& \left(2i\omega c \partial_z + i\omega\dot{\alpha} +
\alpha(\omega^2 - \omega_0^2) \right) {\cal E}
\nonumber\\
&=& 2i\omega\left(c\partial_z{\cal E} +
\frac{\dot{\alpha}}{2}\,{\cal E} \right)
\end{eqnarray}
%%%%%%
when the carrier frequency $\omega$ is equal to the transparency
resonance $\omega_0$. We thus obtain the propagation equation
\cite{Leo:Fleisch}
%%%%%%
\begin{equation}
\label{leo:ee} (1+\alpha)\;\partial_t{\cal E} +
c\;\partial_z{\cal E} + \frac{\dot{\alpha}}{2} {\cal E} = 0 \,.
\end{equation}
%%%%%%
In order to understand the principal behavior of ordinary
slow-light pulses, we consider two cases --- a spatially varying
yet time-independent group index and a spatially uniform yet
time-dependent $\alpha$.

When the group index does not change in time, the propagation
equation (\ref{leo:ee}) has the simple solution
%%%%%%
\begin{equation}
\label{leo:unitime} {\cal E}(t,z) = {\cal E}_0\Big(t -
\int\frac{\d z}{v_g} \Big)\,,
\end{equation}
%%%%%%
in terms of the group velocity (\ref{leo:vg}). At each space
point $z$ the pulse raises and falls in precisely the same way.
However, because light is slowed down, the spatial shape of the
pulse shrinks by a factor of $v_g/c$ compared with the pulse
length in vacuum, for example by $10^{-7}$ for a group velocity of
$30 {\rm m}/{\rm s}$. The intensity of the pulse is unaffected,
despite the enormous pulse shortening, and the pulse energy has
gone into the amplification of the control beam.

Consider the other extreme, a spatially uniform EIT\index{EIT}
medium with adjustable group velocity. In this case, the solution
of the propagation equation (\ref{leo:ee}) is
%%%%%%
\begin{equation}
\label{leo:unispace} {\cal E}(t,z) = {\cal E}_0\Big(z - \int\!
v_g\, \d t\Big)\,\sqrt{v_g/c} \,.
\end{equation}
%%%%%%
The pulse envelope ${\cal E}$ propagates again with the group
velocity $v_g$ but the pulse length is not changed. However, the
spectrum of the pulse around the carrier frequency $\omega_0$
shrinks by a factor of $v_g/c$. Additionally, the intensity drops
by $v_g/c$ as well. The ratio between the control
(\ref{leo:alpha}) and the pulse intensity (\ref{leo:unispace})
remains large,
%%%%%%
\begin{equation}
(1+\alpha)\,\frac{|E_c|^2}{|{\cal E}_0|^2} = \left(|E_c|^2 +
\frac{n_A}{2}\,
\frac{\kappa_{13}^2}{\kappa_{12}^2}\,\frac{\hbar\omega_0}
{\varepsilon_0}\right)\,|{\cal E}_0|^{-2} \,,
\end{equation}
%%%%%%
even in the limit of a vanishing control field when $v_g$
vanishes, as long as $n_A$ is large (for a sufficiently dense
medium). Therefore, the reaction of the probe field is remarkably
consistent with the requirements for undisturbed dark-state
dynamics \cite{Leo:Fleisch}. One can freeze light without losing
control \cite{Leo:Liu,Leo:Philips}.

\index{slow light!pulse|)}
%------------------------------------------------------
\section{Effective field theory}
%------------------------------------------------------

After having studied two examples of pulse propagation in an
EIT\index{EIT} medium, we develop an effective field theory of
slow light. We generalize the wave equation (\ref{leo:wave}) to
three-dimensional space and calibrate the electric field strength
in appropriate units,
%%%%%%
\begin{equation}
E_p(t,{\bf x}) = \left(\frac{\hbar}{\varepsilon_0}\right)^{1/2}
\omega_0\, \varphi(t,{\bf x}) \,.
\end{equation}
%%%%%%
We introduce the Lagrangian
%%%%%%
\begin{equation}
\label{leo:lagrangian0} {\mathscr L} = \frac{\hbar}{2}\left((1 +
\alpha)(\partial_t\varphi)^2 - c^2(\nabla\varphi)^2 -
\alpha\,\omega_0^2\varphi^2\right)
\end{equation}
%%%%%%
and see that the wave equation (\ref{leo:wave}) is the resulting
Euler--Lagrange equation. We have chosen the prefactor of the
Lagrangian (\ref{leo:lagrangian0}) such that ${\mathscr L}$
agrees with the free-field Lagrangian (\ref{leo:l0}) for zero
$\alpha$ and frequencies around $\omega_0$. Therefore we regard
${\mathscr L}$ as the effective Lagrangian of slow light.

Let us use the Lagrangian (\ref{leo:lagrangian0}) to calculate
the energy balance of slow light. According to Noether's theorem
\cite{Leo:Weinberg} we obtain the energy density
%%%%%%
\begin{equation}
I = \frac{\hbar}{2}\left((1 + \alpha)(\partial_t\varphi)^2 +
c^2(\nabla\varphi)^2 + \alpha\,\omega_0^2\varphi^2\right)
\end{equation}
%%%%%%
and the energy flux (Poynting vector)
\begin{equation}
{\bf S} = -\hbar c^2 (\partial_t\varphi)(\nabla \varphi) \,.
\label{leo:energy}
\end{equation}
%%%%%%
As a consequence of the wave equation (\ref{leo:wave}) we obtain
the energy balance
%%%%%%
\begin{equation}
\partial_t I + \nabla \cdot {\bf S} =
\frac{\hbar\dot{\alpha}}{2}\left(\dot{\varphi}^2 +
\omega_0^2\varphi^2\right) \,.
\end{equation}
%%%%%%
Temporal changes in the control field, modifying the group index
(\ref{leo:alpha}), do not conserve energy. In fact, the experiment
\cite{Leo:Liu} indicates that the control beam can amplify light
stored in an EIT\index{EIT} medium with zero group velocity. In
the experiment \cite{Leo:Liu}, slow light enters the
EIT\index{EIT} sample and is then frozen inside by turning off
the control field. Switching on the control releases the stored
light. The pulse emerges with an intensity that depends on the
control field and that may exceed the initial intensity, in
agreement with equation (\ref{leo:unispace}). Clearly, this
phenomenon is only possible if energy is indeed transferred from
the control beam to the probe light.

%------------------------------------------------------
\section{Moving media}
%------------------------------------------------------

An EIT\index{EIT} medium slows down light, because the medium is
extremely dispersive, reacting differently to the different
frequency components of a pulse. The peak position of the pulse
is the place where the frequency components interfere
constructively. By slightly modifying the phase velocity of each
component the medium influences strongly their interference,
slowing down the pulse dramatically.

The extreme spectral sensitivity of slow light can be also
applied to observe optical phenomena in moving media, caused by
the Doppler\index{Doppler effect} effect. A uniformly moving
medium would not present an interesting case, though, because
uniform motion just produces a global frequency shift. However,
slow light is a superb tool in detecting non-uniform motion such
as rotation \cite{Leo:LPGyro}.  To understand the principal
effect of slow light in moving media, we modify the Lagrangian
(\ref{leo:lagrangian0}) to account for the Doppler effect. We
assume that (\ref{leo:lagrangian0}) is valid in frames co--moving
with the medium and transform back to the laboratory frame. We
note that $(\partial_t\varphi)^2 - c^2(\nabla\varphi)^2$ is a
Lorentz invariant and focus on the first--order Doppler effect in
the $\alpha(\partial_t\varphi)^2$ term, assuming the realistic
case of non-relativistic medium velocities. We replace
$\partial_t$ by $\partial_t + {\bf u}\cdot\nabla$ and neglect the
term quadratic in $u$. In this way we obtain the Lagrangian of
slow light in a moving medium
%%%%%%
\begin{equation}
{\mathscr L} = {\mathscr L}_0 - \frac{\alpha}{c^2}\,{\bf S}\cdot
{\bf u} \,,\quad {\bf S} = -\hbar c^2 (\partial_t\varphi)(\nabla
\varphi)
\end{equation}
%%%%%%
in terms of the Lagrangian ${\mathscr L}_0$ for the medium at
rest (\ref{leo:lagrangian0}). We see that the flow ${\bf u}$
couples to the Poynting vector ${\bf S}$ of slow light, similar
to the R\"ontgen interaction of moving dipoles in electromagnetic
fields \cite{Leo:LPRoentgen}. We obtain from ${\mathscr L}$ the
Euler--Lagrange equation
%%%%%%
\begin{equation}
\label{leo:wavemove} \left(\partial_t(1+\alpha)\partial_t - c^2
\nabla^2 + \alpha\,\omega_0^2 + \partial_t\alpha\, {\bf
u}\cdot\nabla + \nabla\cdot \alpha {\bf u}\,\partial_t \right)
\varphi = 0
\end{equation}
%%%%%%
with the differential operators acting on everything to the right.
For frequencies near the EIT\index{EIT} resonance $\omega_0$ we
represent the positive-frequency part $\varphi^{(+)}$ of
$\varphi$ as
%%%%%%
\begin{equation}
\varphi^{(+)} = \varphi_0\,e^{-i\omega_0 t}
\end{equation}
%%%%%%
and perform similar approximations as in Section 5. We obtain the
Schr\"odinger-type equation
%%%%%%
\begin{equation}
\left[i\,\frac{\lbar}{v_g}\,\left(\partial_t -
\frac{\dot{v_g}}{2v_g}\right) + \frac{1}{2}\right]\varphi_0 =
\frac{1}{2}\left(-i\lbar\nabla + \frac{\alpha}{c}\,{\bf u}
\right)^2\varphi_0 - \frac{\alpha^2 u^2}{2c^2}\,\varphi_0
\end{equation}
%%%%%%
with the effective Planck constant reduced by $2\pi$
%%%%%%
\begin{equation}
\lbar = \frac{c}{\omega_0} \,.
\end{equation}
%%%%%%
The flow has a two-fold effect: On one hand, the velocity ${\bf
u}$ appears similar to an effective vector potential, for example
as the magnetic vector potential acting on an electron wave, and,
on the other hand, the hydrodynamic pressure proportional to
$u^2$ acts similarly to an electric potential. A vortex flow will
generate the optical equivalent of the
Aharonov--Bohm\index{Aharonov--Bohm effect} effect, see reference
\cite{Leo:LPSlow} for details.

A moving slow-light medium is also equivalent to an effective
gravitational field \cite{Leo:Papers}. Consider monochromatic
light at exact resonance frequency $\omega_0$. In this case, we
can write the wave equation (\ref{leo:wavemove}) in the form of a
Klein--Gordon equation in a curved space-time,
%%%%%%
\begin{equation}
\partial_\mu \left( f^{\mu\nu} \; \partial_\nu\varphi^{(+)}\right) = 0,
\end{equation}
%%%%%%
with $\partial_\nu = (\partial_t/c,\nabla)$ and
%%%%%%
\begin{equation}
\label{leo:metric} f^{\mu\nu} = \sqrt{-g}\,g^{\mu\nu} = \left[
\begin{array}{cc}
1 & \alpha\,{\bf u}/{c} \\
\alpha\,{\bf u}/{c} & -{\bf 1}
\end{array}
\right].
\end{equation}
%%%%%%
Here $g^{\mu\nu}$ represents the effective space-time metric
experienced by monochromatic slow light in a moving medium, to
first order in $u/c$. We easily find the determinant $g$ of the
inverse of $g^{\mu\nu}$ from the relation
%%%%%%
\begin{equation}
{\rm det} f = -g^2/g = -g =
-\left(1+\alpha^2\,\frac{u^2}{c^2}\right) \,.
\end{equation}
%%%%%%
The effective space-time line element $\d s^2$ is, up to a
conformal factor,
%%%%%%
\begin{equation}
\label{leo:line} \d s^2 = c^2\; \d t^2 + 2\alpha\,\d t\,{\bf
u}\cdot \d{\bf x} - \d{\bf x}^2\,,
\end{equation}
%%%%%%
resembling the line element of a moving coordinate system,
%%%%%%
\begin{equation}
\label{leo:movsys} \d s^2 = (c^2-u^2)\d t^2 + 2\,\d t\,{\bf
u}\cdot \d{\bf x} - \d{\bf x}^2 \,,
\end{equation}
%%%%%%
for example of a rotating system. The parameter $\alpha$
quantifies the degree to which the motion of the medium is
transferred to the propagation of light in the medium, the degree
of dragging. The group index is thus equivalent to Fresnel's
dragging coefficient \cite{Leo:LPGyro,Leo:LPstor}. For slow light
$\alpha$ is very large indeed. Therefore, slow light is able to
sense minute flow variations. Even subtle quantum flows imprint
phase shifts onto slow light that are detectable using
phase-contrast microscopy \cite{Leo:Hecht}.

Sound waves in a fluid experience the flow as an effective
space-time metric as well \cite{Leo:Visser}. The acoustical line
element is proportional to the element (\ref{leo:movsys}) of a
moving system of coordinates, with two crucial differences: The
flow is not subject to the rigidity of moving coordinate systems
and, more importantly, in the acoustic metric the speed of light,
$c$, is replaced by the speed of sound. A supersonic flow
surpasses the sound barrier and can, under suitable
circumstances, generate an artificial event horizon where the
flow speed $u$ reaches $c$. Here it is necessary that the $g_{00}$
element of the metric vanishes. In the slow-light metric
(\ref{leo:line}) the all-important term $-u^2 \d t^2$ is missing
in $g_{00}$, at least to the level of approximation we are
considering here. We obtain a term proportional to $-u^2 \d t^2$
when we include effects of higher-order Doppler\index{Doppler
detuning} detuning. The critical velocity is then of the order of
$c/\sqrt{\alpha}$. To observe the quantum effects of light
generated by a horizon we would need to employ a steep profile of
the flow speed. This causes a severe problem, because the
Doppler\index{Doppler detuning} detuning will exceed the
transparency window of EIT\index{EIT}. The Doppler\index{Doppler
effect} effect plays a beneficial role in the sensitivity of slow
light to motion, but it will also cause significant light
absorption when one attempts to reach an artificial\index{analog
horizon} event horizon. The medium will certainly turn black, but
not into a black\index{optical black hole} hole. However, one
could employ a spatially varying profile of the group index to
create an interface that resembles an event horizon for slow
light and that avoids this problem \cite{Leo:Catastrophy}. Slow
light offers a variety of options for interesting experiments
exploiting the analogs of light in media with effects in other
areas of physics, and new ideas continue to emerge.

%------------------------------------------------------
\section{Summary}
%------------------------------------------------------

Light has been slowed down dramatically \cite{Leo:Slow} or even
stopped completely \cite{Leo:Liu,Leo:Philips}. To understand how
this feat has been achieved, we studied the physics behind slow
light, starting from basic first principles of the light-matter
interaction. We first turned to ordinary optical media, so as to
later contrast them with slow-light media based on
Electromagnetically-Induced Transparency. We studied slow light
in two regimes --- in a stationary regime both dominated and
limited by relaxation and the control light, and in a dynamic
regime almost free from dissipation. Then we analyzed the typical
behavior of slow-light pulses, before developing an effective
field theory of slow light that we have subsequently generalized
to moving media.

It is certainly amazing how much a clever combination of atomic
physics and optics can achieve, but it is also important to
understand the principal limits of the techniques applied. These
limits depend on the details of the physics behind the scene. We
have tried to elucidate the details of slow light without going
into too much detail, using models that are simple, but not too
simple.

\index{slow light|)}
%------------------------------------------------------

%------------------------------------------------------
\end{document}